\documentclass[a4paper,superscriptaddress,aps,prl,twocolumn,floatfix,citeautoscript]{revtex4-2}
\usepackage[utf8]{inputenc}
\usepackage{amsmath}
\usepackage{graphicx}
\usepackage{cleveref}
\usepackage{xcolor}
\usepackage{cancel}
\usepackage{booktabs}
\usepackage{ulem}

\usepackage{feynmp-auto}
\newcommand{\lcpq}{Laboratoire de Chimie et Physique Quantiques, Universit\'e de Toulouse, UPS, CNRS, and European Theoretical Spectroscopy Facility (ETSF), 118 route de Narbonne, F-31062 Toulouse, France}
\newcommand{\lpt}{Laboratoire de Physique Th\'eorique, CNRS, Universit\'e de Toulouse, UPS, and European Theoretical Spectroscopy Facility (ETSF), 118 route de Narbonne, F-31062 Toulouse, France}

\begin{document}
\title{The multi-channel Dyson equation: coupling many-body Green's functions}

\author{Gabriele Riva}
\affiliation{\lcpq}
\email{griva@irsamc.ups-tlse.fr}
\author{Pina Romaniello}
\affiliation{\lpt}
\email{pina.romaniello@irsamc.ups-tlse.fr}
\author{J. Arjan Berger}
\email{arjan.berger@irsamc.ups-tlse.fr}
\affiliation{\lcpq}

\begin{abstract}
We present the multi-channel Dyson equation that combines two or more many-body Green's functions to describe the electronic structure of materials.
In this work we use it to model photoemission spectra by coupling the one-body Green's function with the three-body Green's function.
We demonstrate that, unlike methods using only the one-body Green's function, our approach puts the description of quasiparticles and satellites on an equal footing.
We propose a multi-channel self-energy that is static and only contains the bare Coulomb interaction, making frequency convolutions and self-consistency unnecessary.
Despite its simplicity, we demonstrate with a diagrammatic analysis that the physics it describes is extremely rich.
Finally, we present a framework based on an effective Hamiltonian that can be solved for any many-body system using standard numerical tools.
We illustrate our approach by applying it to the Hubbard dimer and show that it is exact both at 1/4 and 1/2 filling.
\end{abstract}

\maketitle

The Dyson equation is a fundamental equation of many-body perturbation theory.
This integral equation links a bare propagator to the corresponding interacting propagator through a kernel that contains the many-body effects. 
For example the non-interacting one-body Green's function is linked to the interacting one through the self-energy.
Similar Dyson-like equations exist for higher-order propagators, e.g., the Bethe-Salpeter equation, which is related to the two-body Green's function.
Each of these Dyson equations involves just a single type of many-body Green's function.
In this work we extend the concept of the Dyson equation to a multi-channel version which allows us to couple two or more Green's functions. 
Thus, for a given physical process, the pertinent physics of each of the Green's functions involved can be exploited.
We focus here on the coupling of the one-body Green's function and the three-body Green's function to calculate photoemission spectra, but our method can be easily generalized to the coupling of other many-body Green's functions.

To model photoemission spectra most methods use the one-body Green's function (1-GF) $G_1$ as the fundamental quantity, since, within the sudden approximation, its spectral function $A_1(\omega) = |\textrm{Im} [G_1(\omega)]|/\pi$ can be easily linked to photoemission spectra.
Among those methods, the $GW$ approach has been by far the method of choice for most applications.
Unfortunately, the $GW$ approach has several shortcomings both at the fundamental and practical level: 
1) $GW$ quasiparticle energies are not accurate for strongly correlated systems~\cite{Dis21-2,Dis23-1};
2) $GW$ satellites are, in general, not accurate even in systems that are not strongly correlated~\cite{Guz11};
3) fully self-consistent $GW$ calculations are not routine, in particular for solids, although it can be done~\cite{Kut10,Kut16,Kut17,Gru18,Yeh_2022};
4) partial self-consistent $GW$ calculations, which are numerically simpler~\cite{Fal04,VanS06,Kot07,Bru06},
exhibit problems such as discontinuities in potential energy surfaces~\cite{Loo18,Ver18,Ber21};
5) self-consistent methods such as $GW$, can converge to an unphysical solution~\cite{Lan12,Ber14,Sta15,Tar17}.

Therefore, much effort has been devoted to explore novel routes to calculate accurate photoemission spectra. 
All these efforts have mainly focused on improving the one-body Green’s function or the corresponding self-energy, albeit  from different perspectives~\cite{Guz11,Lan12,Ste14,Dis16,Back22,Van22,Dis23-1,Rom09-1,Dis21-1,Riv22}. 
Some results are promising, but a unified theory for the accurate description of photoemission spectra at both weak and strong correlation is still missing.
In this context, we have recently proposed to use the three-body Green's function (3-GF) $G_3$ as the fundamental quantity of the theory~\cite{Riv22}.
To be precise, the fundamental quantity is the hole-hole-electron plus the electron-electron-hole contributions to $G_3$. 
With some notable exceptions (see, e.g., Refs.~\cite{Sch04,Bar01,Bar07,Man94,Mar99,Dei16,Tor19,Pav21}), the 3-GF or equivalent three-body objects have, so far, not been studied much in the literature.
%So far, the 3-GF has mainly been used to obtain approximations to the one-body self energy~\cite{Sch04,Bar01,Bar07,Pav21,Mar99,Man94} or for the description of trion states~\cite{Dei16,Tor19}.
One of the main advantages of our approach is that, unlike methods based on $G_1$, such as $GW$, it puts quasiparticles and satellites on an equal footing. This is particularly important in the strongly correlated limit in which the quasiparticle picture breaks down since quasiparticles and satellites mix. 
We have given a proof of principle of our approach using the exactly solvable Hubbard dimer.~\cite{Riv22} 

We now want to make our approach applicable to real systems. 
To achieve this, we introduce three new developments in this article: 
1) we introduce a multi-channel Dyson equation that couples the one-particle channel with the three-particle channel;
2) we propose an approximation to the corresponding multi-channel self-energy that can be applied to any many-body system; 
3) we give a general framework based on an effective Hamiltonian that can be solved using standard numerical tools. 
Our multi-channel self-energy is static and only contains the bare Coulomb interaction, i.e., it is not a functional of $G_3$.
As a consequence of these features, neither frequency convolutions nor self-consistency is required in our approach.
Despite the simplicity of this self-energy, the physics it describes is rich thanks to the mixing of $G_3$ into $G_1$. 
We will show that in this way screening beyond the random-phase approximation (RPA) as well as ladder diagrams are introduced. 
Finally we will illustrate our approach by applying it to the Hubbard dimer at 1/4 and 1/2 filling and show that it is exact in both cases. 

Our starting point is the following Dyson equation for $G_3$~\cite{Riv22},
\begin{equation}\label{Dyson:eq}
    G_{3}(\omega)=G_{03}(\omega)+G_{03}(\omega) \Sigma_{3}(\omega)G_{3}(\omega),
\end{equation}
where $G_{03}(\omega)$ is the noninteracting 3-GF and $\Sigma_3$ is the three-body self-energy. 
We project this equation in the basis set of one-electron spinorbitals $\{\phi_i\}$ that diagonalizes $G_{03}$.
In this basis $G_{03}$ is naturally partitioned in a one-particle channel $G_{01}$, which corresponds to the noninteracting 1-GF, 
and a three-particle channel $G_{03}^{3\text p}$, which corresponds to the two-electrons-one-hole ($2e1h$) and two-holes-one-electron ($2h1e$) contributions to $G_{03}$. 
We can thus rewrite $G_{03}$ in the following matrix representation
\begin{equation}%\label{G0matrix:eq}
    G_{03}(\omega)=\begin{pmatrix}
        G_{01}(\omega) & 0 \\
        0 & G_{03}^{3\text p}(\omega)
    \end{pmatrix},
    \label{Eqn:G03}
\end{equation}
where
\begin{align}\label{G0convstat:eq}
    G_{01(im)}(\omega)&=\frac{\delta_{im}}{\omega-\epsilon^0_i+i\eta\text{sign}(\epsilon^0_i-\mu)}, \\ 
    %G^0_{im}&=i\gamma^0_{im}=i\delta_{im}f_i  \label{static:eq}\\
    G_{03(i>jl;m>ok)}^{3 \text p}(\omega)&=
    %[G^0_{im}G^0_{jo}G^0_{lk}](\omega)=
    \frac{\delta_{im}\delta_{jo}\delta_{lk}(f_i-f_l)(f_j-f_l)}{\omega-\epsilon^0_i-(\epsilon^0_j-\epsilon^0_l)+i\eta\text{sign}(\epsilon^0_i-\mu)},
    \label{convG0:eq}
\end{align}
in which $\epsilon^0_i$ are the orbital energies of the non-interacting system, $f_i$ are the corresponding occupation numbers ($f_i=1$ for $\epsilon_i\leq \mu$ and 0 otherwise), and $\mu$ is the chemical potential.   
Equations (\ref{G0convstat:eq}) and (\ref{convG0:eq}) set the space in which Eq.~(\ref{Dyson:eq}) is solved.
The restrictions $i>j$ and $m>o$ in Eq.~\eqref{convG0:eq} avoid double counting while the occupation numbers $(f_i-f_l) (f_j-f_l)$ restricts $G_{03}$ to its $2h1e$ and $2e1h$ contributions.
Without loss of generality, the non-interacting 3-GF can be replaced with an independent-particle 3-GF.
It is convenient to use a Hartree-Fock 3-GF since it already contains Hartree and exchange contributions.
In this way, the three-body self-energy in Eq.~\eqref{Dyson:eq} is reduced to its correlation part.
Therefore, in the following, $G_{03}$ will denote the Hartree-Fock 3-GF.

With the partition of $G_{03}$ given in Eq.~\eqref{Eqn:G03}, Eq.~\eqref{Dyson:eq} becomes a multichannel Dyson equation, in which the multi-channel self-energy is defined as
\begin{align}\label{selfmatrix:eq}
    \Sigma_3=\begin{pmatrix}
        \Sigma^{1 \text p} & \Sigma^{\text c}\\
        \tilde \Sigma^{\text c} & \Sigma^{3\text p}
    \end{pmatrix}.
\end{align}
For practical calculations we need an approximation to $\Sigma_3$. 
To achieve this we correlate only pairs of particles in $\Sigma^{3 \text p}$. We let each pair interact via a direct and an exchange interaction, i.e., at the RPA+exchange (RPAx) level, as in the electron-hole and particle-particle channels of the 2-GF~\cite{Oni02,Stri88,Kad64}. %~\cite{Bar01,Bar07,Pav21}
The four-point couplings $\Sigma^{\text c}$ and $\tilde{\Sigma}^{\text c}$ correspond to two-particle channels which, for consistency, are also treated at the RPAx level.
Since all the correlation is included in $\Sigma^{\text c}$, $\tilde \Sigma^{\text c}$, and $\Sigma^{3\text p}$ the head, $\Sigma^{1\text p}$, vanishes.

We thus arrive at the following static approximation,
\begin{align}
    %\Sigma^{1p}_{i;m}&=0    \label{self1p:eq}    \\
    \Sigma^{3 \text p}_{ijl;mok}&=\!\! [(1\!-\!f_i)\!(1\!-\!f_j)f_l\!-\!f_if_j(1\!-\!f_l)]\nonumber \\ &\times [(1-f_i)(1-f_m)+f_if_m][\delta_{lk} \bar v_{ijom} \nonumber \\ &+\!\delta_{mj}\bar v_{iklo} \!+\! \delta_{io} \bar v_{jklm} \!-\! \delta_{oj}  \bar v_{iklm} \!-\! \delta_{im}  \bar v_{jklo}] ,\label{selfthird:eq}\\
    \Sigma^{\text c}_{i;mok}&=\bar v_{ikom}, \label{selfcoupling:eq} \\
     \tilde \Sigma^{\text c}_{ijl;m}&=\bar v_{ijlm}, \label{selfcouplingtilde:eq}\\
%     & \Sigma^{3p}_{ijl;mok}=\begin{pmatrix}
%        \Sigma_{\tilde c \tilde c' \tilde v;cc'v}^{2p/1h} & 0 \\
%        0 & \Sigma_{\tilde v \tilde v' \tilde c;vv'c}^{2h/1p} \end{pmatrix}  \label{selfthirdmatrix:eq}
    \Sigma^{1 \text p}_{i;m}&=0, \label{selfone:eq}
\end{align}
where $\bar v_{ikom}=v_{ikom}-v_{ikmo}$ with
\begin{equation}
    v_{ikom}=\int dx_1 dx_2 \phi^*_i(x_1)\phi^*_k(x_2)v(\mathbf{r}_1,\mathbf{r}_2)\phi_o(x_2)\phi_m(x_1).\label{potential:eq}
\end{equation}
The first term on the right-hand side of Eq.~\eqref{selfthird:eq}, i.e, the one involving $\delta_{lk}\bar v_{ijom}$, accounts for all the two-particle interactions in $\Sigma^{3\text p}$.
%, while the last four terms account for all the one-electron-one-hole interactions.
The last four terms account for all the one-electron-one-hole interactions, since there are two electron-hole couples and for each of these couples either the electron or the hole can exchange with the third particle.
The occupation numbers in Eq.~\eqref{selfthird:eq} ensure that $\Sigma^{3 \text p}$ is block diagonal and it has opposite signs for the $2e1h$ channel and the $2h1e$ channel.
We note that our approximate three-body self-energy is hermitian.
The approximations in Eqs.~\eqref{selfthird:eq}-\eqref{selfcouplingtilde:eq} can also be obtained using a technique similar to the adiabatic diagrammatic construction~\cite{Sch78,Sch83,Nie84,Bin21}.

It is instructive to represent the multi-channel Dyson equation in Eq.~\eqref{Dyson:eq} diagrammatically according to

\begin{widetext}
    \begin{equation}\label{matrixdiagram:eq}
    \begin{gathered}
        \includegraphics[width=0.9\textwidth,clip=]{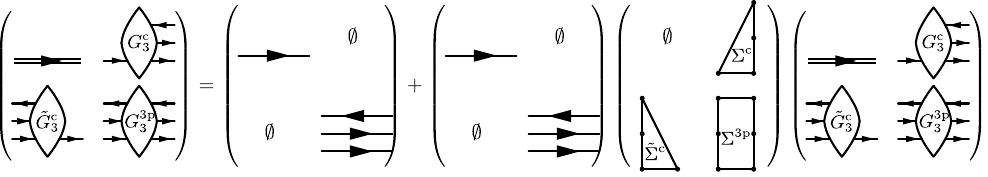}
    \end{gathered}
    \end{equation}
From Eq.~(\ref{matrixdiagram:eq}) we see that the 3-GF contains the 1-GF (\raisebox{-0.3\totalheight}{\includegraphics[width=0.1\textwidth]{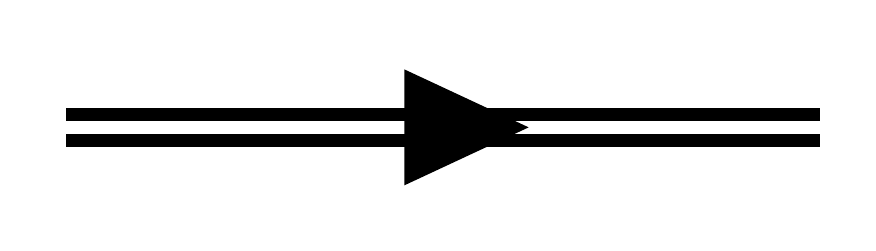}})
plus an explicit three-body part $G_3^{3p}$ (\raisebox{-0.3\totalheight}{\includegraphics[width=0.03\textwidth]{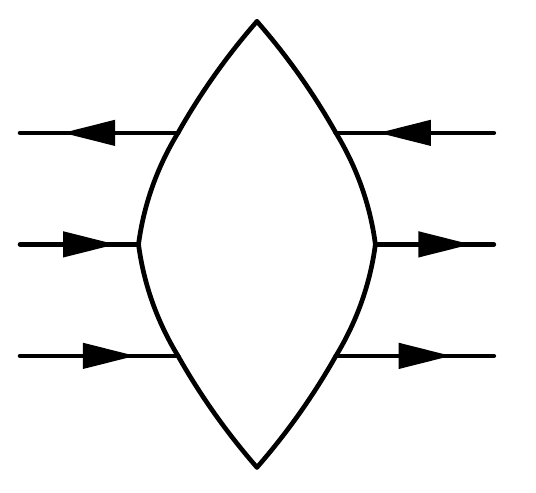}})
and the couplings between the 1-GF and $G_3^{3p}$, namely $G^c_3$ (\raisebox{-0.3\totalheight}{\includegraphics[width=0.03\textwidth]{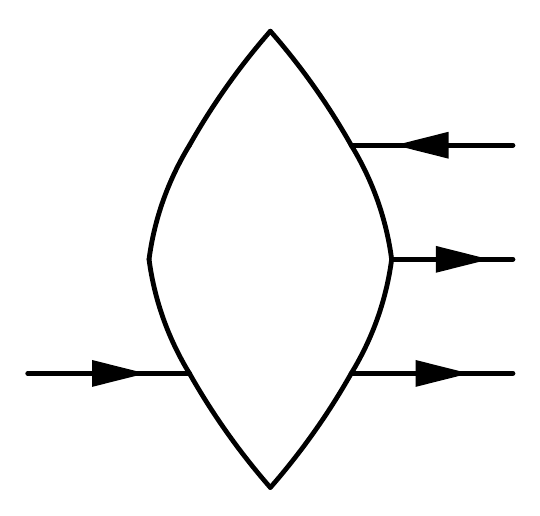}}) and $\tilde{G}^c_3$ (\raisebox{-0.3\totalheight}{\includegraphics[width=0.03\textwidth]{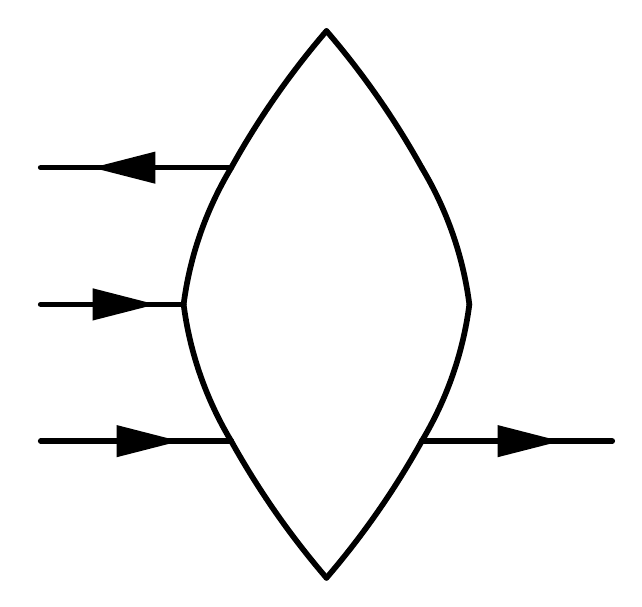}}).
The $G_3^{3p}$, $G^c_3$ and $\tilde{G}^c_3$ terms add correlation to the non-interacting 1-GF (\raisebox{-0.1\totalheight}{\includegraphics[width=0.06\textwidth]{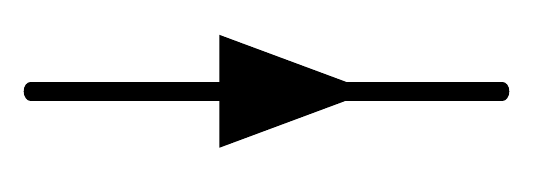}}) and 3-GF (\raisebox{-0.4\totalheight}{\includegraphics[width=0.05\textwidth]{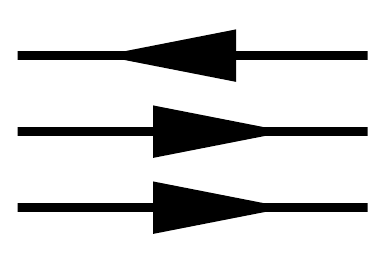}}).
We represent $\Sigma^{c}_{ijl,m}$ and $\tilde{\Sigma}^c_{i,mok}$ by right triangles to reflect their dimensions.

It is convenient to represent the multi-channel self-energy in real space. 
Details about the change of basis are given in the supplemental material~\cite{supmat}. 
The $2e1h$ channel of Eq.~\eqref{selfthird:eq} is given by
\begin{equation}
    \begin{gathered}
            \includegraphics[width=0.92\textwidth,clip=]{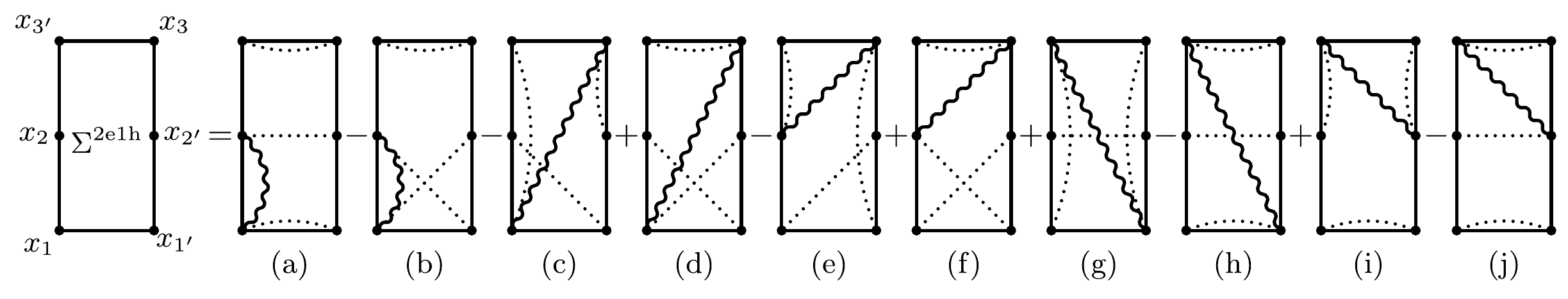} 
    \end{gathered}
    \label{Eqn:Dyson_diag}
\end{equation}
\end{widetext}
The $2h1e$ diagrams are equal to the $2e1h$ diagrams in Eq.~\eqref{Eqn:Dyson_diag} but with an overall minus sign.
In the diagrams given above, each dotted line
represents a Dirac delta, merging the two points it connects, and each wiggly line represents the bare Coulomb interaction. 
The coupling terms are given by
\begin{align}
    \begin{gathered}
        \includegraphics[width=0.38\textwidth,clip=]{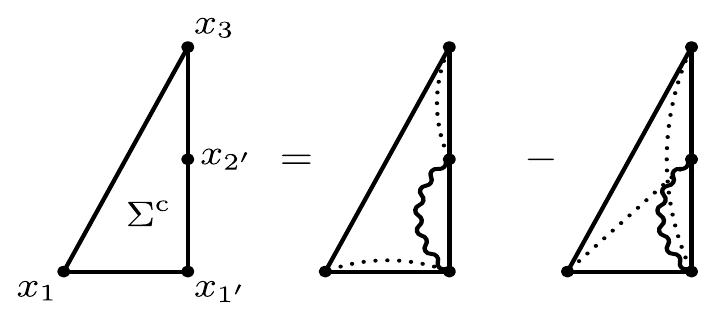}
    \end{gathered}
        \label{Eqn:coupling1} \\   
    \begin{gathered}
        \includegraphics[width=0.38\textwidth,clip=]{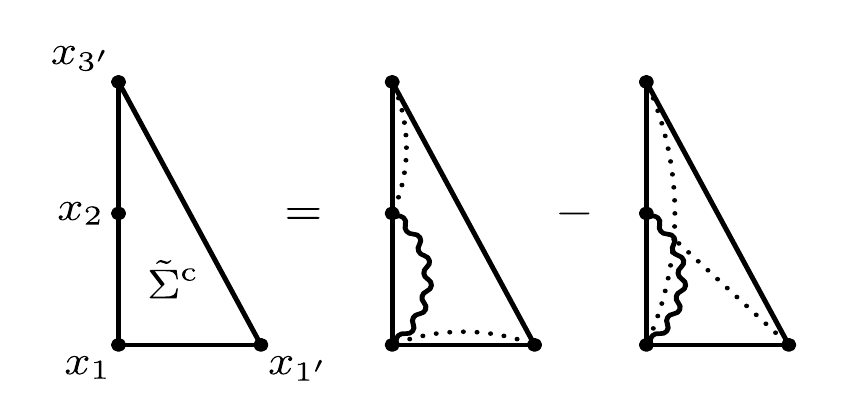}
    \end{gathered}      
    \label{Eqn:coupling2}              
\end{align}

which correspond to Eq.~\eqref{selfcoupling:eq}, and Eq.~\eqref{selfcouplingtilde:eq}, respectively.

To understand which diagrams of $G_1$ are included at each order in our approach, it suffices to iterate the multi-channel Dyson equation in Eq.~\eqref{matrixdiagram:eq} and inspect the head of the matrix.
One iteration does not change the head, i.e., no correlation is added to $G_{01}$.
A second and third iteration yield the following two self-energy insertions, respectively,
\begin{align}
\begin{gathered}
        \hspace*{-0.5cm}
       \includegraphics[width=0.2\textwidth,clip=]{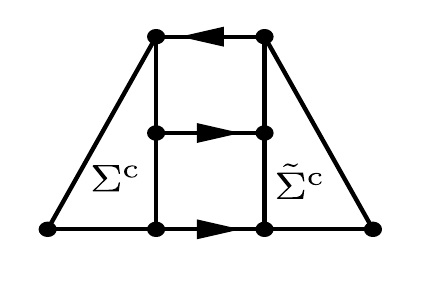}  \hspace*{-0.5cm}
    \includegraphics[width=0.25\textwidth,clip=]{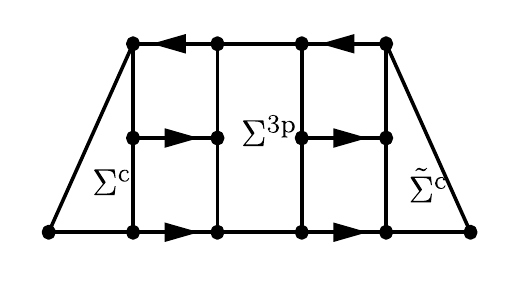}     
\end{gathered}\label{Eqn:diagrams23}
\end{align}

In general, the $n$-th order diagram is obtained by inserting between $\Sigma^c$ and $\tilde\Sigma^c$ in the $(n-1)$th-order diagram
a $\Sigma^{3\text p}$ rectangle linked to $G_{03}^{3\text p}$.
Inserting Eqs.~\eqref{Eqn:coupling1} and \eqref{Eqn:coupling2} into the diagram on the left in \eqref{Eqn:diagrams23} yields both second-order proper skeleton diagrams,
\begin{align}
    \begin{gathered}
                   \includegraphics[width=0.4\textwidth,clip=]{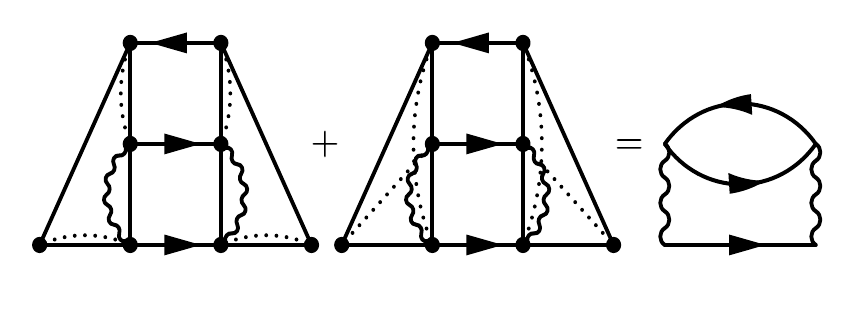} 
    \end{gathered}
    \\
    \begin{gathered}
                   \includegraphics[width=0.42\textwidth,clip=]{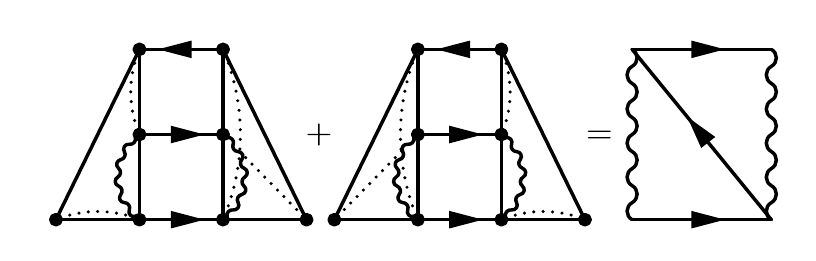}  
    \end{gathered}
\end{align}

We note that each diagram on the right-hand side of Eqs.~(16) and (17) is the sum of two diagrams. 
Because of the restriction of the space mentioned before, both diagrams are needed, and there is no double counting.
It can be verified that our approach is complete up to second order~\cite{Mathar_2007}.
The details of the diagrammatic analysis are explained in the supplemental material~\cite{supmat}. 
It can be verified that inserting Eqs.~\eqref{Eqn:Dyson_diag}-\eqref{Eqn:coupling2} into the diagram on the right in Eq.~\eqref{Eqn:diagrams23} yields all ten third-order proper skeleton diagrams, which include both bubble and ladder diagrams.
For example, a bubble diagram is obtained as follows,

\begin{equation}
    \begin{gathered}
            \includegraphics[width=0.42\textwidth,clip=]{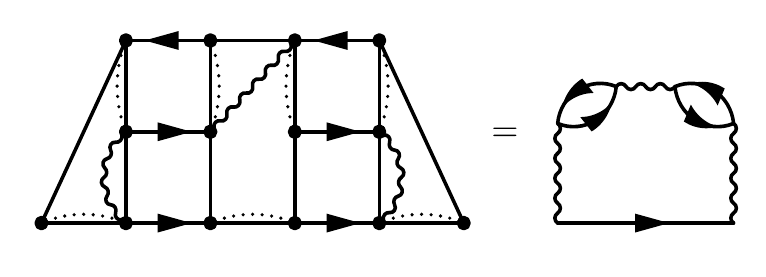}  
    \end{gathered}
    \label{Thirdorderbuble}
\end{equation}

and a ladder diagram as

\begin{equation}
    \begin{gathered}
        \includegraphics[width=0.42\textwidth,clip=]{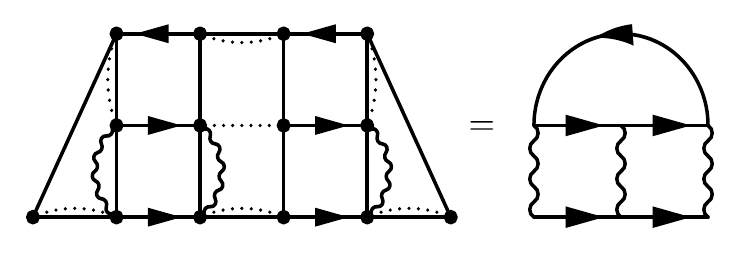}         
    \end{gathered}
\end{equation}

The above bubble diagram is also contained in the $GW$ approximation which contains bubble diagrams up to infinite order. 
In our approach, higher-order diagrams are obtained by iterating the multi-channel Dyson equation \eqref{matrixdiagram:eq} further.
By doing so, it is possible to check that all $GW$ diagrams are included in our approach, and screening effects are thus accounted for. 
Moreover, our approach goes beyond the RPA screening included in $GW$.  
As an example, we report the following fourth-order diagram
\begin{equation}
    \begin{gathered}
            \includegraphics[width=0.42\textwidth,clip=]{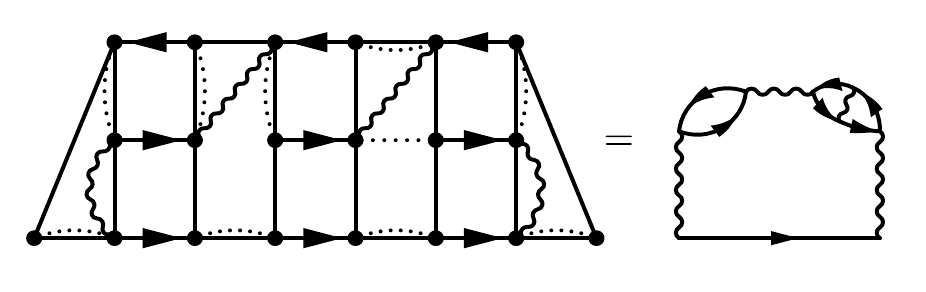}  
    \end{gathered}
\end{equation}

where a first-order vertex correction to the RPA bubble is shown. 
% In practice, we solve the multi-channel Dyson equation in Eq.\eqref{Dyson:eq}. 
% As a consequence, contributions to all orders in the interaction are included.
Although the self-energy only involves the bare Coulomb potential, screening beyond the RPA is naturally included in the multi-channel Dyson equation.
Therefore, there is no need to use a screened interaction in our theory since this would lead to the double counting of diagrams.

The total number of proper skeleton diagrams included in Eq.~\eqref{Dyson:eq}, 
within the approximate multi-channel self-energy in Eqs.\eqref{selfthird:eq}-\eqref{selfone:eq}, at order $n$ is given by $\frac{10^{n-2}}{2^{n-3}}$ for $n\ge 2$. 
We note that one could include more correlation by dressing $G^{3 \text p}_{03}$ beyond HF using, e.g., second Born, $GW$ or the T-matrix within the quasiparticle approximation.
\footnote{In this case one should assume that the correlated $G^{3 \text p}_{03}$ is diagonal in the same basis as $G_{03}$.
This approximation is also used, for example, to derive the cumulant approximation \cite{Guz11,martin_reining_ceperley_2016}.}.
Finally, the frequency-independent $\Sigma_3$ used here, corresponds to a frequency-dependant $\Sigma_1(\omega)$.
The frequency dependence is included through $G^{3 \text p}_{03}(\omega)$.

To obtain an equation that can be solved in practice using standard numerical tools, we use
a strategy similar to the one used for the Bethe-Salpeter equation \cite{Oni02}, i.e., we map Eq.~\eqref{Dyson:eq} onto an effective three-particle equation according to

\begin{equation}\label{G3effective:eq}
    G_{3(ijl;mok)}=\left[%\begin{pmatrix}
     %   H^{1-B}_{i'm'} & H^c_{i'm'o'k'}\\
    %    \tilde H^c_{i'j'l'm'} & H^{3-B}_{i'j'l';m'o'k'}
   % \end{pmatrix}
   \omega I-H^{\text{eff}}\right]^{-1}_{ijl;mok}
\end{equation}
in which the effective Hamiltonian $H^{\text{eff}}$ is given by
\begin{align}\label{Heff:eq}
    H^{\text{eff}}=\begin{pmatrix}
        H^{1 \text p} & \Sigma^{\text c} \\
        \tilde {\Sigma}^{\text c} & H^{3 \text p}
    \end{pmatrix},
\end{align}
where 
\begin{align}
    H^{1 \text p}_{i;m}&=\epsilon^0_i\delta_{im}\label{H1p:eq},\\
    H^{3\text p}_{ijl;mok}&=(\epsilon^0_i-(\epsilon^0_l-\epsilon^0_j))\delta_{im}\delta_{jo}\delta_{lk}+\Sigma^{3 \text p}_{ijl;mok}. \label{H3p:eq}
\end{align}
Since our final goal is to calculate photoemission spectra, which are linked to the one-body-channel of $G_3$, it suffices to extract the head from $G_3$.

A straightforward diagonalisation of Eq.~\eqref{Heff:eq} would scale as $N^9$, where $N$ is the number of electrons in the system. However, the scaling can be reduced to $N^6$ using standard iterative methods such as the Haydock-Lanczos solver~\cite{Hay72,Schm03,Her05}. We also note that methods that describe trions have a similar scaling as our approach, and they have successfully been applied to real systems~\cite{Dei16,Tor19}.

\begin{figure}
    \centering
    \includegraphics[width=\columnwidth]{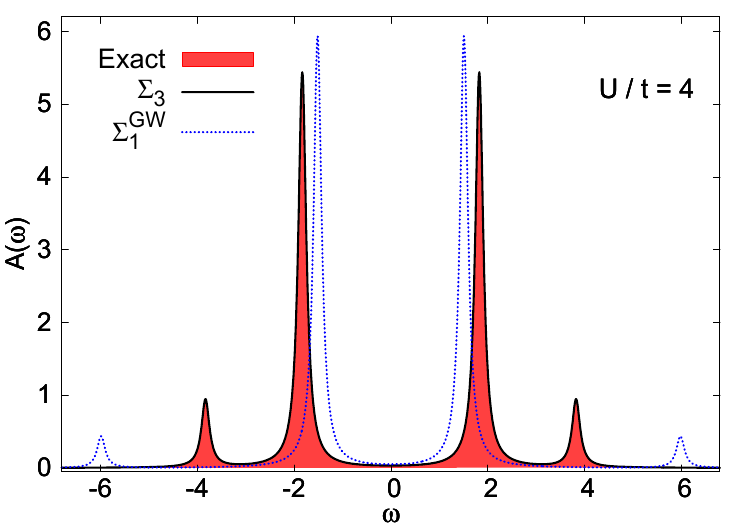}
    \caption{The spectral function of the Hubbard dimer at 1/2 filling obtained at vari
    ous levels of theory. Exact result (red filled); the multi-channel Dyson equation (black solid line); the $GW$ approximation (blue dotted line).}
    \label{Hubbard1/2:fig}
\end{figure}

Finally, as an illustration, we apply our approach on the symmetric Hubbard dimer at $1/4$ and $1/2$ filling for $U/t=4$.
The details of the calculations can be found in the supplemental material~\cite{supmat}.
In Figs.~\ref{Hubbard1/4:fig} and \ref{Hubbard1/2:fig}, we report the spectral functions we obtain with the multi-channel Dyson equation and compare it to the $GW$ spectral function.
While the $GW$ spectrum fails to describe the satellites, our approach yields exact result for both quasiparticles and satellites.
We note that with our approximation to $\Sigma_3$ we also obtain the exact $G_3$ for the Hubbard dimer.

In conclusion, we presented a strategy in which two or more $n$-body Green's functions are coupled using a multi-channel Dyson equation. 
Here we applied this strategy to the coupling of $G_1$ and $G_3$ to put quasiparticles and satellites on an equal footing.
We demonstrated formally that we thus go beyond standard approximations, such as $GW$.
%We showed that we thus obtain all diagrams up to third order in the interaction as well as many diagrams beyond third order.
We showed that, contrary to $GW$, our approach yields exact results for the spectral function of the symmetric Hubbard dimer at one-fourth and one-half filling.
By treating quasiparticles and satellites on an equal footing, our approach is particularly suited to describe strong correlation, which is exemplified by the exact results we obtain for the Hubbard dimer. Therefore, in future work, we will use it to improve the \textit{ab initio} description of strongly correlated materials.
The concept of a multi-channel Dyson equation is general and can be applied also to other couplings. 
For example one can couple $G_2$ and $G_4$ to get an accurate description of double excitations or bi-excitons. Work in this direction is in progress.
\begin{figure}
    \centering
    \includegraphics[width=\columnwidth]{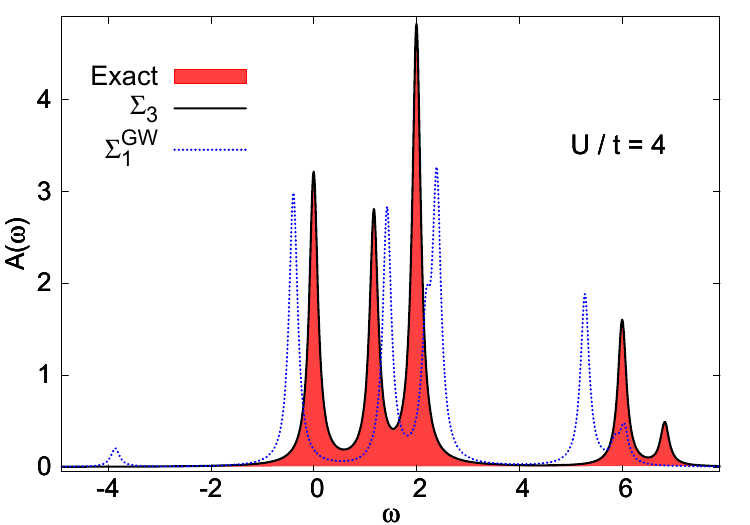}
    \caption{The spectral function of the Hubbard dimer at 1/4 filling obtained at various levels of theory. Exact result (red filled); the multi-channel Dyson equation (black solid line); the $GW$ approximation (blue dotted line). We note that the $GW$ spectrum has two unphysical peaks at $\omega\simeq -4$ and $\omega \simeq 5.9$ which are caused by the self-screening problem $GW$ suffers from~\cite{Rom09-1}.}
    \label{Hubbard1/4:fig}
\end{figure}

\textit{Acknowledgment:}
%We thank Carlo Barbieri and the late Peter Schuck for useful discusssions.
We thank the French “Agence Nationale de la Recherche (ANR)” for financial support (Grant Agreements No. ANR-19-CE30-0011 and No. ANR-22-CE30-0027).

%\bibliographystyle{ieeetr} 
%\bibliography{main} % Entries are in the refs.bib file


\end{document}